\documentclass[aps,twocolumn,preprintnumbers,floatfix,showpacs,prl]{revtex4-1}

\usepackage{amsmath,amssymb}
\usepackage{dcolumn}
\usepackage{braket}
\usepackage{dsfont}
\usepackage{feynmp}
\usepackage{hhline}
\usepackage{graphicx}
\usepackage{xfrac}
\usepackage{paralist}
\usepackage{color}

\definecolor{olive}{rgb}{0.0,0.4,0.0}
\definecolor{darkred}{rgb}{0.9,0,0}
\newcommand{\bB}{\bf \color{blue}}
\newcommand{\bR}{\bf \color{darkred}}
\newcommand{\bG}{\bf \color{olive}}
\begin{document}

\title{A Generalization of Non-Abelian Anyons in Three Dimensions}
\author{Sagar Vijay}
\author{Liang Fu}
\affiliation{Department of Physics, Massachusetts Institute of Technology,
Cambridge, MA 02139, USA}
\begin{abstract}
We introduce both an exactly solvable model and a coupled-layer construction for an exotic, three-dimensional phase of matter with immobile topological excitations that carry a protected internal degeneracy.  Unitary transformations on this degenerate Hilbert space may be implemented by braiding certain point-like excitations. 
This provides a new way of extending non-Abelian statistics to three-dimensions.
\end{abstract}
\maketitle

A core concept in the quantum theory of indistinguishable particles is quantum statistics. While fundamental particles in nature only obey either Bose or Fermi statistics, the notion of particle statistics also applies to quasiparticles -- emergent, point-like excitations in many-body systems with an energy gap -- through the geometric phase accumulated in a braiding process, whereby two identical quasiparticles are adiabatically interchanged.
The allowed statistics are constrained by the topology of quasiparticle  trajectories in the braiding process.  As a consequence of the non-trivial topology of braids in $(2+1)$-dimensional spacetime, certain quasiparticles in two dimensions -- known as anyons \cite{wilczek} -- are allowed to have statistics other than Bose or Fermi \cite{statistics,Halperin}.  Non-Abelian anyons  \cite{MooreRead, Kitaev} are particularly interesting, since a state of well-separated non-Abelian anyons carries a degeneracy that cannot be lifted through local perturbations. Braiding a pair of these excitations can implement a unitary transformation on this space of states. Efforts to search for non-Abelian anyons are underway \cite{Read, Stern}. 

Do particles with neither Bose nor Fermi statistics exist in three dimensions? A standard argument rules out this possibility based on the observation that exchanging a pair of particles twice in a (3+1)-dimensional spacetime is topologically equivalent to no exchange. This implies that two exchanges must leave the quantum state invariant, hence a single exchange can only generate a phase factor $\pm 1$, corresponding to Bose or Fermi particle statistics, respectively.

Despite this no-go argument, the possibility of lifting anyons to three dimensions has long fascinated physicists. It is known that in three-dimensional lattice gauge theories with a discrete, non-Abelian gauge group, point-like charge excitations can carry a protected internal degeneracy with integer ``quantum dimension'' \cite{XGW, Deligne}.  Nonetheless, these excitations still have Bose or Fermi statistics, and their internal state remains unchanged under braiding.  Other possibilities have also been explored. An intriguing study \cite{TeoKane} suggested that Majorana zero modes on the surface of a superconducting topological insulator \cite{FuKane} display braiding properties analogous to non-Abelian anyons, despite living in a three-dimensional system. However, these Majorana zero modes cannot be spatially separated at finite energy cost \cite{Nayak}. Such objects, now commonly referred to ``twist defects'', are {\it not} deconfined quasiparticles.

 \begin{figure}
 \includegraphics[trim = 0 0 0 0, clip = true, width=0.5\textwidth, angle = 0.]{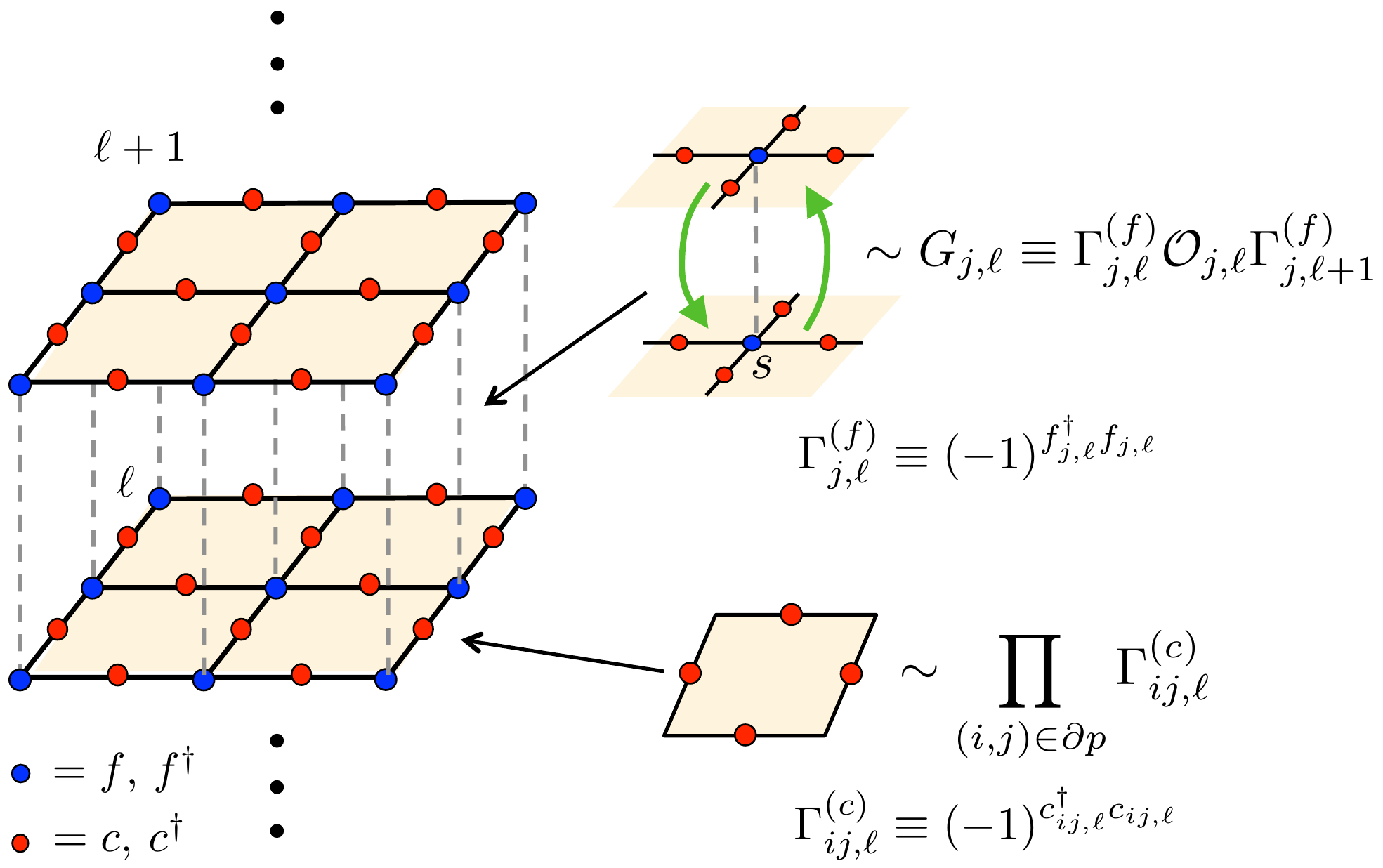}
 \caption{{\bf The Model:} We consider layers of of complex fermions on the sites of a square lattice ($f$), coupled to fermions that lie on the links ($c$) which play the role of a static $Z_{2}$ gauge field.  Within each layer, the hopping and pairing interaction of the fermions are mediated by the fermions on the links, which interact at each plaquette as shown.   The gauge symmetry of the Hamiltonian (\ref{eq:H}) is implemented by the operator $G_{j,\ell} = \Gamma_{j, \ell -1}^{(f)}\,\mathcal{O}_{j,\ell}\,\Gamma_{j,\ell+1}^{(f)}$ which couples adjacent layers, with $\mathcal{O}_{j,\ell}$ as defined in the main text. The ground-state realizes an exotic phase with immobile, point-like excitations that carry a protected internal degeneracy.}
  \label{fig:Lattice}
\end{figure}

In this Letter, we demonstrate the existence of a new type of deconfined point-like excitation in three dimensions, which carries a protected degeneracy as non-Abelian anyons in two dimensions do, but cannot move freely without paying an energy cost. These fundamentally immobile particles -- termed ``fractons" \cite{Vijay1} -- have been theoretically discovered in recent years \cite{Haah,Chamon,BLT,Vijay2} and have attracted increasing interest in such diverse fields as topological quantum matter, lattice gauge theory, quantum information, and many-body localization \cite{Vijay2, Dominic,Yoshida, Vijay3,HermeleXie,Hsieh,Rahul,YMLu,Pretko}. Being unable to move, fractons evade the standard argument for particle statistics relying on the topology of worldlines. This enables our generalization of the notion of non-Abelian particles to three dimensions. The non-Abelian fractons found in this work can provide an alternate platform for quantum computation, with increased robustness against thermal errors. 

We introduce two lattice models which support such ``non-Abelian fracton" excitations.  The first is an exactly solvable fermion model that describes a chiral phase where the fractons have {\it irrational} quantum dimension. This phase is constructed from  two-dimensional layers of $p_{x}+ip_{y}$ superconductors interacting with $Z_{2}$ gauge field, and by further coupling these layers in a nontrivial way. The interlayer coupling turns fluxes into deconfined, immobile point-like excitations, i.e., fractons.
As a result, when the parent superconducting state on each layer is topologically nontrivial, the fluxes acquire a $\sqrt{2}$ quantum dimension, yielding ``non-Abelian" fracton excitations in a three-dimensional phase.  Certain pairs of these fractons behave as non-Abelian anyons with well-defined statistics.  The second model we present is constructed from interacting layers of two-dimensional $G$-gauge theories for a finite group $G$ \cite{Quantum_Double}, which are stacked along all three directions. This isotropic layer construction yields point-like, immobile excitations with integer quantum dimension.
In both models, fractons can only be created, in multiples of four, at the corners of an operator with support on a membrane-like region. This defining property of fractons is fundamentally distinct from that of anyons, which are created in pairs, at the ends of string-like Wilson line operators.

We begin with a detailed description of our first model.  Consider a vertical stack of two-dimensional square lattice planes, as shown in Fig. \ref{fig:Lattice}.  Within each plane $\ell$, there are two types of fermions living on the sites (denoted $f_{j,\ell}$) and bonds  (denoted $c_{ij,\ell}$) of the lattice. Our solvable Hamiltonian is given by:
\begin{align}\label{eq:H}
H= \sum_{\ell} H_\ell - K\sum_{\ell, p} B_{p,\ell} - \sum_{\ell, j} G_{j,\ell}.
\end{align}
where $H_{\ell}$ is defined for every layer ($\ell$), $B_{p,\ell}$ for every plaquette ($p$), and $G_{j,\ell}$ at every site $j$ on adjacent planes $\ell$ and $\ell+1$, as indicated in Fig. \ref{fig:Lattice}.

The first term in the Hamiltonian takes the form
\begin{align}
H_{\ell} = &\sum_{\langle i,\,j \rangle} \left[-t \, f^{\dagger}_{i,\ell}\,\Gamma^{(c)}_{ij,\ell}\,f_{j,\ell} + \Delta_{ij}\,f^{\dagger}_{i,\ell}\,\Gamma^{(c)}_{ij,\ell}f^{\dagger}_{j,\ell} + \mathrm{h.c.}\right] \nonumber\\
&- \mu\sum_{i}f^{\dagger}_{i,\ell}f_{i,\ell}.
\end{align}
where
\begin{align}
\Gamma^{(c)}_{ij,\ell} \equiv & (-1)^{c^{\dagger}_{ij,\ell}c_{ij,\ell}} = 1- 2c^{\dagger}_{ij,\ell}c_{ij,\ell}
\end{align}
is the parity of the fermions along link $(i,j)$ in layer $\ell$.  $H_{\ell}$ describes $f$ fermions in a $(p_{x}+ip_{y})$-wave paired state on each plane, with a nearest-neighbor hopping $t$ and pairing $\Delta_{ij}$ on the two-dimensional square lattice, with $\Delta_{ij}=\Delta$ on $+x$ links and $i \Delta$ on $+y$ links. Importantly, these ``matter fermions'' $f$ interact with ``gauge fermions'' $c$ via Ising gauge coupling, with the parity of the $c$ fermions on the links $\Gamma^{(c)} = \pm 1$ playing the role of a $Z_2$ gauge field.

In the second term of (\ref{eq:H}), the operator $B_{p,\ell}$ is given by the product of the fermion parities $\Gamma_{ij,\ell}^{(c)}$ along the links surrounding plaquette $p$ in layer $\ell$
\begin{align}
B_{p,\ell} \equiv \prod_{\langle i,\,j\rangle\in\partial p}\Gamma^{(c)}_{ij,\ell}
\end{align}
The  $\pm 1$ eigenvalue of $B_{p, \ell}$ measures the $Z_{2}$ flux of the gauge field through this plaquette.

The last term in the Hamiltonian describes an interaction between fermions on adjacent layers.  The operator $G_{j,\ell}$, which is shown schematically in Fig. \ref{fig:Lattice}, is defined on the sites of the square lattice as
\begin{align}\label{eq:gauge}
G_{j,\ell} \equiv \Gamma_{j,\ell}^{(f)}\,\mathcal{O}_{j,\ell}\,\Gamma_{j,\ell+1}^{(f)}
\end{align}
where $\Gamma_{j, \ell}^{(f)} = (-1)^{f^{\dagger}_{j,\ell}f_{j,\ell}}$ is the fermion parity of the matter fermion on site $s$ in layer $\ell$ and
\begin{align}
\mathcal{O}_{j,\ell} = \prod_{i\in\mathrm{star}(j)}\left[ \frac{1}{2}(c^{\dagger}_{ij,\ell} - c_{ij,\ell})(c^{\dagger}_{ij,\ell+1} + c_{ij,\ell+1})\right]\nonumber
\end{align}
is an eight-body interaction that couples the fermions on the links forming a ``star" configuration around site $s$ in layer $\ell$ and in layer $\ell+1$. The operators $G_{i,\ell}$  mutually commute, and satisfy $G_{i, \ell}^2 = 1$.
 Remarkably, $G_{i,\ell}$
also commute with the Hamiltonian (\ref{eq:H}) since (i) $H_{\ell}$ is invariant under the local $Z_2$ transformation $f_{i, \ell} \rightarrow - f_{i, \ell}$, $\Gamma^{(c)}_{ij,\ell}\rightarrow - \Gamma^{(c)}_{ij,\ell}$ and (ii) $B_{p, \ell}$  overlaps with $G_{i, \ell}$ on two links.
For $t,\mu, \Delta, K \ll 1$, all low-lying eigenstates of $H$ satisfy the ``gauge constraint'' 
\begin{align}\label{Gauss}
G_{i,\ell}\ket{\Psi} = \ket{\Psi}
\end{align}
at every site $i$ and layer $\ell$.
In the following,  we will restrict our attention to these gauge-invariant states.

It is instructive to first study the model (\ref{eq:H}) in the limit $t = \Delta = 0$. The ground state is then simply a direct product state of the gauge and matter fermions
$\ket{\Psi_{\mathrm{gs}}} = \ket{g_c} \otimes \ket{0_f}$.
Here $\ket{0_f}$ is the vacuum state  of matter fermions with $\Gamma^{(f)}=1$ at every lattice site (assuming $\mu>0$), while $\ket{g_c}$ is the ground state of the reduced Hamiltonian for the gauge  fermions
\begin{align}\label{Hc}
H_{c} = - K \sum_{\ell, p}B_{p,\ell} - \sum_{j,\ell}\mathcal{O}_{j,\ell}.
\end{align}
This commuting Hamiltonian was introduced in Ref. \cite{Vijay1} as an exactly solvable model for a fracton topological phase, whose universal properties are robust under any local perturbations. 
When placed on the $L \times L \times L$ three-torus, $H_c$ exhibits $2^{3L-3}$ degenerate ground states that are locally indistinguishable.  
An elementary $\pi$-flux excitation is obtained when the eigenvalue of an operator $B_{p, \ell}$ on a plaquette is flipped. Remarkably, these $\pi$-flux excitations can
only be created in multiples of four by acting with a {\it membrane} operator on the ground state, which flips the eigenvalues of $B_{p, \ell}$'s at the corners of the membrane. Therefore, a single $\pi$-flux is a fracton---it cannot be moved without creating additional fractionalized excitations.  
A self-contained discussion of this model is provided in the supplemental material \cite{Supp_Mat}.

Apart from the $\pi$-flux, our model also hosts gapped fermionic quasiparticles originating from the matter fermions on every layer.  The fermionic excitations carry gauge charge, however, and must bind additional excitations in order to be gauge invariant.  Adding or removing a bare matter fermion $f_{j, \ell}$ flips the parity of $\Gamma_{j,\ell}^{(f)}$ and locally violates the gauge constraint (\ref{Gauss}).  Instead, a gauge-invariant quasiparticle is obtained by binding a matter fermion with an  excitation of gauge fermions having $\mathcal{O}_{j,\ell} = -1$ and $\mathcal{O}_{j,\ell-1} = -1$.
Such a gauge excitation can only move within the plane, and has $\pi$ mutual statistics with the $\pi$-flux excitation, which is a fracton.  As a result, the fermionic quasiparticle also has $\pi$ mutual statistics with the $\pi$-flux, reminiscent of a conventional $Z_2$ gauge theory with charged matter fields.

When $t, \Delta \ne 0$, gauge-invariant eigenstates of $H$ take the general form $\ket{\mathrm{\Psi}} =  P \ket{ \eta } \otimes\ket{\varphi}_\eta$ where  $\ket{\eta}$ is a state of gauge fermions with a fixed fermion parity on every link, so that $\Gamma^{(c)}_{ij,\ell} \ket{\eta}  =\eta_{ij, \ell} \ket{ \eta}$ with $\eta_{ij, \ell} = \pm 1$.
$\ket{\varphi}_\eta$ is an eigenstate of the matter fermions in gauge field configuration $\eta$.  That is, $\ket{\varphi}_\eta$ is obtained by substituting the operators $\Gamma^{(c)}_{ij,\ell}$ with their eigenvalues $\eta_{ij,\ell}$ in the first term of $H$, and then solving the resulting quadatric Hamiltonian for the matter fermions. Finally, $P$ is a projection operator, which projects the wavefunction of the matter and gauge fermions onto the gauge invariant subspace
\begin{align}\label{eq:gs}
P \equiv  \prod_{j,\ell}\left(\frac{1 +  G_{j,\ell}}{2}\right).
\end{align}

To verify that the above wavefunction is an eigenstate of $H$, we note that (i) $[P,H]=0$ and that (ii) by construction, $\ket{ \eta } \otimes\ket{\varphi}_\eta$ is an eigenstate of the first two terms in $H$, whose eigenvalue we denote $E(\eta, \varphi)$. It follows that
\begin{align}
H \ket{\mathrm{\Psi}} = & H P \ket{ \eta } \otimes\ket{\varphi}_\eta = P H \ket{ \eta } \otimes\ket{\varphi}_\eta \nonumber \\
= & P \Big( E(\eta, \varphi) - \sum_{j, \ell} G_{j, \ell} \Big) \ket{\eta} \otimes\ket{\varphi}_\eta \nonumber \\
= & \Big[ E(\eta, \varphi) - N \Big] \ket{\mathrm{\Psi}},
\end{align}
where we have used the identity $P G_{j, \ell} = P$, and
$N$ is the number of $f$ fermion sites.  The energy spectrum of $H$ is thus determined by that of the quadratic Hamiltonian for the matter fermions in a fixed gauge flux configuration.



When $K \gg t, \Delta$, the ground state belongs to the zero-flux gauge sector (e.g. with $\eta_{ij,\ell}=1$ on all links), and the matter fermions realize a $p_{x}+ip_{y}$ superconductor on every layer. As  a result of the gauge-matter coupling, $\pi$-flux excitations of the gauge field are now bound to vortices of the $p_x + i p_y$ superconductor.
When $|\mu | > 4t$, the $p_{x}+ip_{y}$ superconducting state of the matter fermions is fully gapped and 
adiabatically connected to the $t=\Delta=0$ limit. Therefore, $\pi$-flux excitations are topologically equivalent to the fractons of the Hamiltonian (\ref{Hc}) of gauge fermions only.  When $|\mu |< 4t$, however, the matter fermions in the zero-flux gauge sector  realize a fully-gapped $p_{x}+ip_{y}$ topological superconductor on every layer. In this case, $\pi$-flux excitations 
become non-Abelian fractons with internal topological degeneracy, as we now show.

Recall that a two-dimensional $p_x + i p_y$ topological superconductor hosts localized Majorana zero modes in vortex cores \cite{Read_Green, Ivanov}. When this superconductor is coupled to a ${Z}_{2}$ gauge field, the $\pi$-flux-vortex composite object becomes a deconfined quasiparticle, which is a well-studied example of a non-Abelian anyon---the Ising anyon \cite{Kitaev}. $N$ well-separated Ising anyons carry $d^N$ degenerate internal states which cannot be split by local perturbations, with the quantum dimension $d = \sqrt{2}$ originating from the Majorana zero mode. 
Bringing two Ising anyons ($\sigma$) together, however, can split the degeneracy, resulting in either a fermionic excitation ($\psi$) or a trivial boson ($1$). This behavior is captured  by the fusion rules
\begin{align}\label{eq:fusion}
 \sigma \times \sigma = 1 + \psi
\end{align}
as well as the rules $\sigma \times \psi = \sigma$, $\,
 \psi \times \psi = 1$ and $1 \times a = a$, with $a = 1, \psi, \sigma$.
Braiding Ising anyons generates a non-trivial unitary transformation on their internal states.

In our three-dimensional model, when $|\mu| < 4t$, the $\pi$-flux bound to a vortex on a single layer hosts a Majorana zero mode, and hence acquires a protected internal degeneracy with quantum dimension $d=\sqrt{2}$. Unlike Ising anyons, however, these $\pi$-fluxes are fundamentally immobile point-like excitations; this property originates from the fractons in the model (\ref{Hc}) for the gauge fermions before  coupling to matter fields. We thus refer to these fractons, with a topologically protected internal degeneracy, as ``non-Abelian fractons''.     

We now study the nature of composite excitations made from a pair of non-Abelian fractons in our model.
First, we show that 
the internal degeneracy of pairs of non-Abelian fractons in \emph{distinct} layers is protected, even when they are brought close together.
Splitting the degeneracy requires quasiparticle tunneling between the two Majorana zero modes.  This process is forbidden, however, since there is no gauge-invariant operator that can transfer quasiparticles between distinct layers.  This can be seen by observing that the local gauge constraint $G_{j,\ell}=+1$ gives rise to conserved fermion parity on every pair of adjacent planes: $\prod_{j}G_{j,\ell} = U^{(f)}_{\ell}U^{(f)}_{\ell+1} = +1$, where $U^{(f)}_{\ell}\equiv \prod_{j}\Gamma^{(f)}_{j,\ell}$ is the fermion parity in layer $\ell$.
This parity conservation naturally forbids inter-layer quasiparticle tunneling in the gauge-invariant subspace, so that a pair of non-Abelian fractons in distinct layers $\ell$ and $\ell'$ -- referred to schematically as $\sigma_{\ell}\times \sigma_{\ell'}$ -- forms a topological excitation with quantum dimension $(d_{\sigma})^2 = 2$.  This excitation is a non-Abelian anyon that can only move within the $xy$ plane. Braiding this non-Abelian anyon around a fracton enclosed in its plane of motion can implement a unitary transformations on this degenerate Hilbert space; a particular example is explicitly given in the supplemental material \cite{Supp_Mat}. 

 \begin{figure}
 $\begin{array}{cc}
 \includegraphics[trim = 0 0 0 0, clip = true, width=0.23\textwidth, angle = 0.]{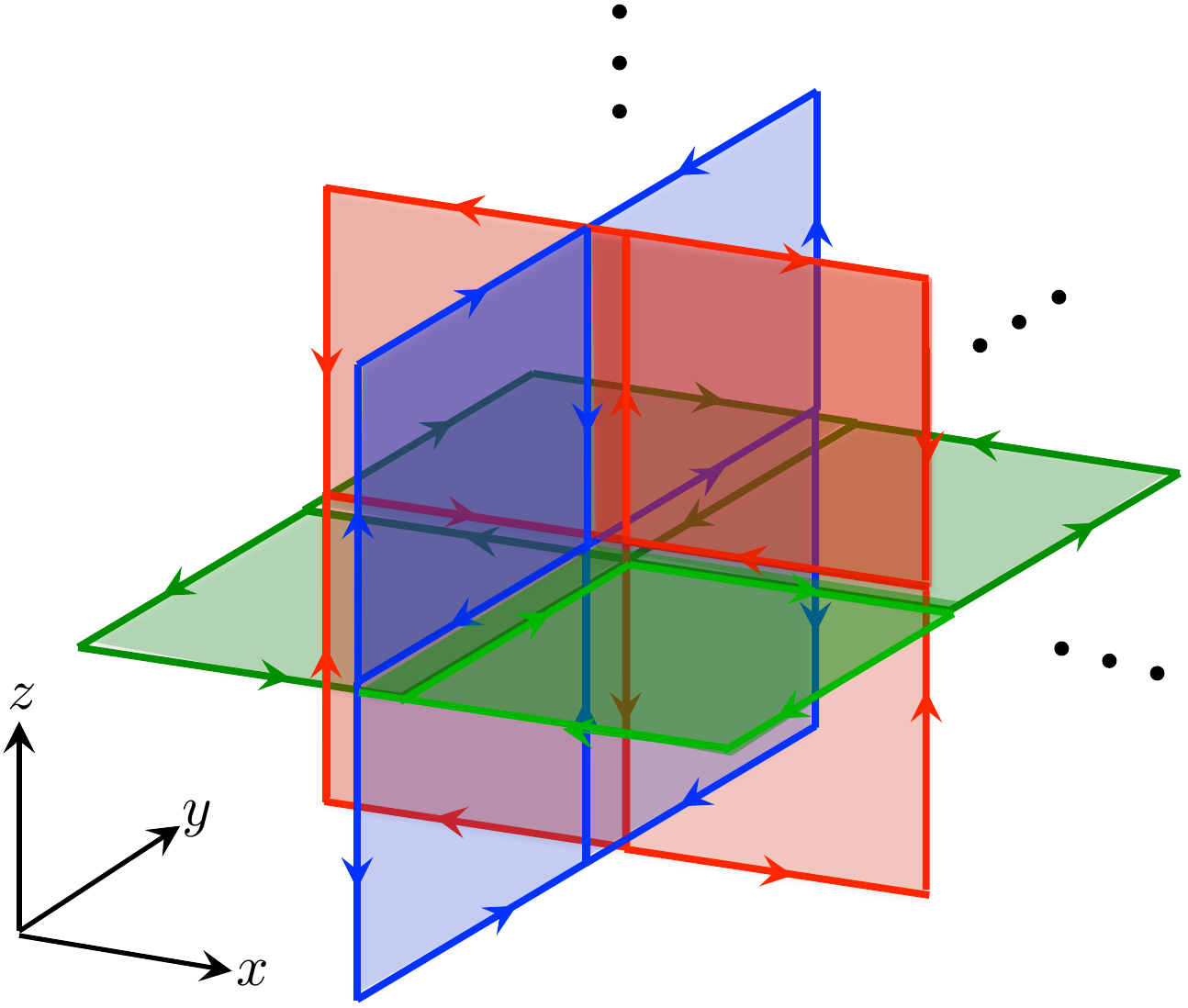} &
 \includegraphics[trim = 0 0 0 0, clip = true, width=0.2\textwidth, angle = 0.]{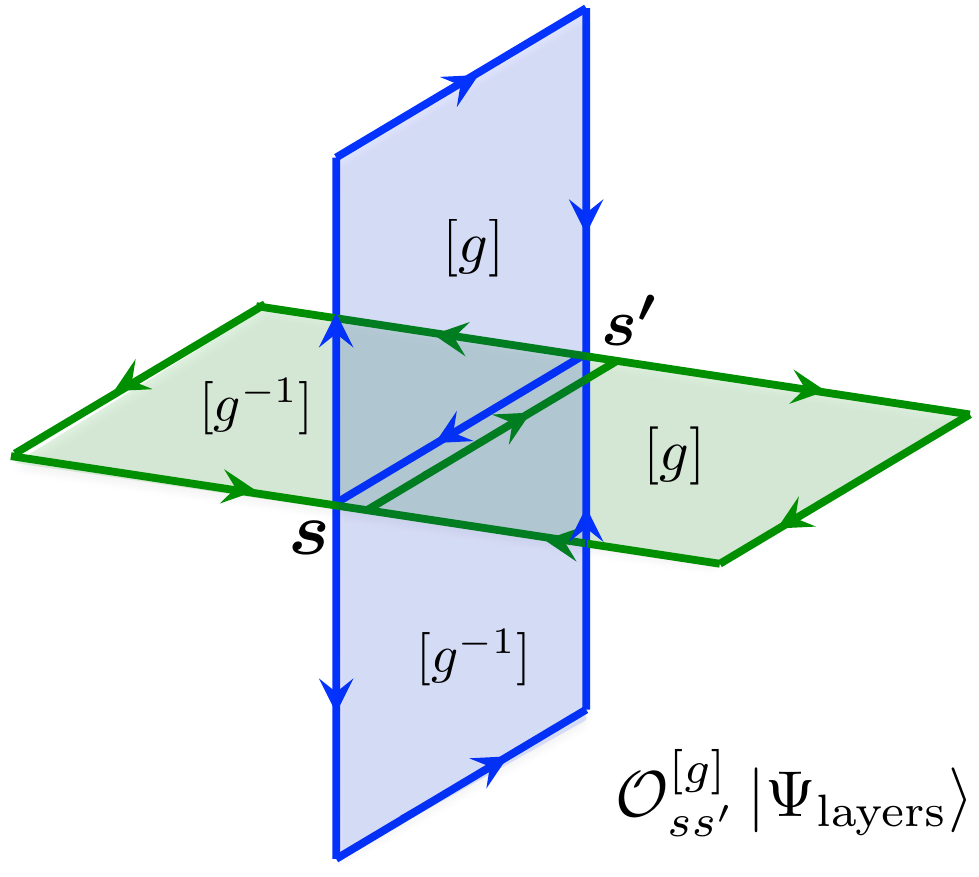}\\
 \text{(a)} & \text{(b)}\\
 &
  \end{array}$\\
  \includegraphics[trim = 0 0 0 0, clip = true, width=0.48\textwidth, angle = 0.]{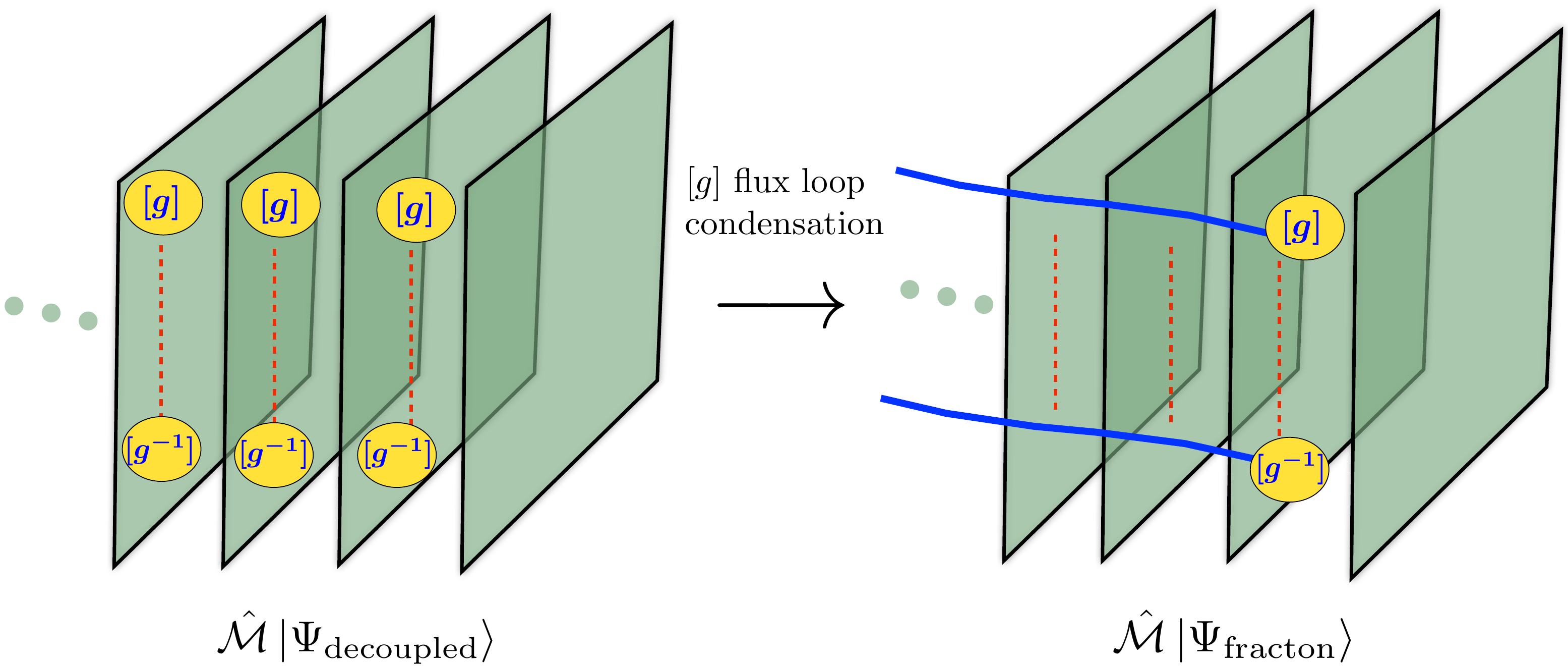}\\
(c)
 \caption{{\bf Coupled 2D $G$ gauge theories:} Intersecting layers of 2D gauge theories for finite the group $G$, stacked in the $xy$, $yz$ and $xz$ planes as shown in (a).  ``Condensing" the excitation in (b) -- a composite of four $[g]$ fluxes -- can produce a fracton topological phase, where the fracton excitation inherits certain properties from the non-Abelian $[g]$ flux as shown in (c) and described in the main text.}
  \label{fig:Quantum_Double_Layers}
\end{figure}

We now summarize an alternate prescription for obtaining a non-Abelian fracton topological phase where the fracton excitations carry integer quantum dimension, which is motivated by an isotropic, layer construction \cite{Vijay3, HermeleXie} of a particular fracton phase introduced in Ref. \cite{Vijay2}.  The starting point for this construction is an array of two-dimensional lattice gauge theories for a finite gauge group $G$ that are stacked in the $xy$, $yz$ and $xz$ directions, as shown in Fig. \ref{fig:Quantum_Double_Layers}a.  Initially, distinct layers are decoupled and each layer is in the deconfined phase of the $G$ gauge theory, which hosts gapped charge and flux excitations that are labeled by irreducible representations and conjugacy classes of $G$, respectively.  Starting from this isotropic array, we now introduce an interaction on the bonds of the lattice; when acting on the ground-state of the decoupled layers, this coupling has the effect of creating a pair of fluxes (labeled $[g]$) in orthogonal layers as shown in Fig. \ref{fig:Quantum_Double_Layers}b. We refer to this as a ``$[g]$ composite flux loop", since this excitation forms a closed loop on the dual lattice.  Increasing the strength of the inter-layer coupling eventually condenses this  excitation, yielding a ground-state which is given by a superposition of loop-like excitations built from the 2D $[g]$ fluxes. 

We observe that a fracton topological phase is obtained by considering the excitations that remain deconfined after this condensation procedure.  When the layers are decoupled, we may act with an array of Wilson line operators in parallel layers to create a sequence of $[g]$ fluxes, as shown in Fig. \ref{fig:Quantum_Double_Layers}c. This excitation carries an $O(L)$ energy cost, where $L$ is the linear dimension of the array.  After condensation of the composite $[g]$ flux loop, the bulk of this excitation is indistinguishable from a configuration of fluxes that appear in the ground-state. As a result, in the condensed phase, the bulk of this excitation costs no energy.  The membrane-like operator formed from the array of Wilson lines may create excitations at its \emph{corners}, however, as the corners appear to be points where the composite $[g]$ flux loops have broken open, as shown in Fig. \ref{fig:Quantum_Double_Layers}c.  Due to the geometry of this operator, these point-like excitations cannot be moved without nucleating other excitations in the system, and we conclude that a single $[g]$-flux has become an immobile fracton excitation.  The charges and dyons of the two-dimensional $G$ gauge theory will generically be confined due to their non-trivial statistics with the $[g]$ flux.  However, the condensation procedure may bind these into emergent excitations with reduced mobility.

When the gauge group $G \cong {Z}_{2}$, this condensation procedure yields the so-called X-cube fracton topological phase \cite{Vijay2, Vijay3, HermeleXie}; $Z_{N}$ generalizations of the X-cube phase may be constructed in a similar fashion \cite{Vijay3}.  However, $G$ can also be a finite  \emph{non-Abelian} group. In this case, condensing a composite flux loop made of non-Abelian fluxes of the $G$ gauge theory will produce a non-Abelian fracton topological phase where the immobile fracton excitations carry a protected internal degeneracy, so long as a single $[g]$-flux remains deconfined after this condensation procedure.  Due to the non-trivial statistics and fusion rules of the $[g]$ fluxes, the general conditions under which this condensation procedure will yield a non-Abelian fracton topological phase is not known, though certain examples may be explicitly analyzed.

In the supplemental material \cite{Supp_Mat}, we study coupled layers of the $S_{3}$ gauge theory, where $S_{3}$ is the permutation group on three elements.  Condensing the composite flux loop formed from the $S_{3}$ fluxes with quantum dimension $3$ yields a fracton topological phase, where the non-Abelian flux becomes an immobile topological excitation.  The condensation procedure also has the effect of binding certain two-dimensional charges into excitations with restricted mobility, while all mobile charge excitations are confined.  For example, while an isolated charge excitation that corresponds to the alternating representation of $S_{3}$ is confined, \emph{pairs} of these charges remain well-defined excitations that may only move along lines.  It would be interesting if a similar layer construction using other ``string-net" models \cite{Levin_Wen} could also yield exotic, 3D phases with immobile fractionalized excitations. 

\acknowledgments
\emph{Acknowledgments:} This work was supported by DOE Office of Basic Energy Sciences,  Division  of  Materials  Sciences  and  Engineering under Award de-sc0010526.  LF is supported partly by  the  David  and  Lucile  Packard Foundation.

\appendix
\section{Checkerboard Model}
In this section, we review the properties of the checkerboard model, which was introduced \cite{Vijay1} as a solvable model for a fracton topological phase with local fermionic excitations.  As described in the main text, the model is recovered in the limit $t = \Delta = 0$ of the Hamiltonian (\ref{eq:H}) as the reduced effective Hamiltonian for the gauge fermions ($c_{ij,\ell}$, $c_{ij,\ell}^{\dagger}$), as explicitly presented in (\ref{Hc}).  

 \begin{figure}
 $\begin{array}{c}
 \includegraphics[trim = 300 200 320 160, clip = true, width=0.3\textwidth, angle = 0.]{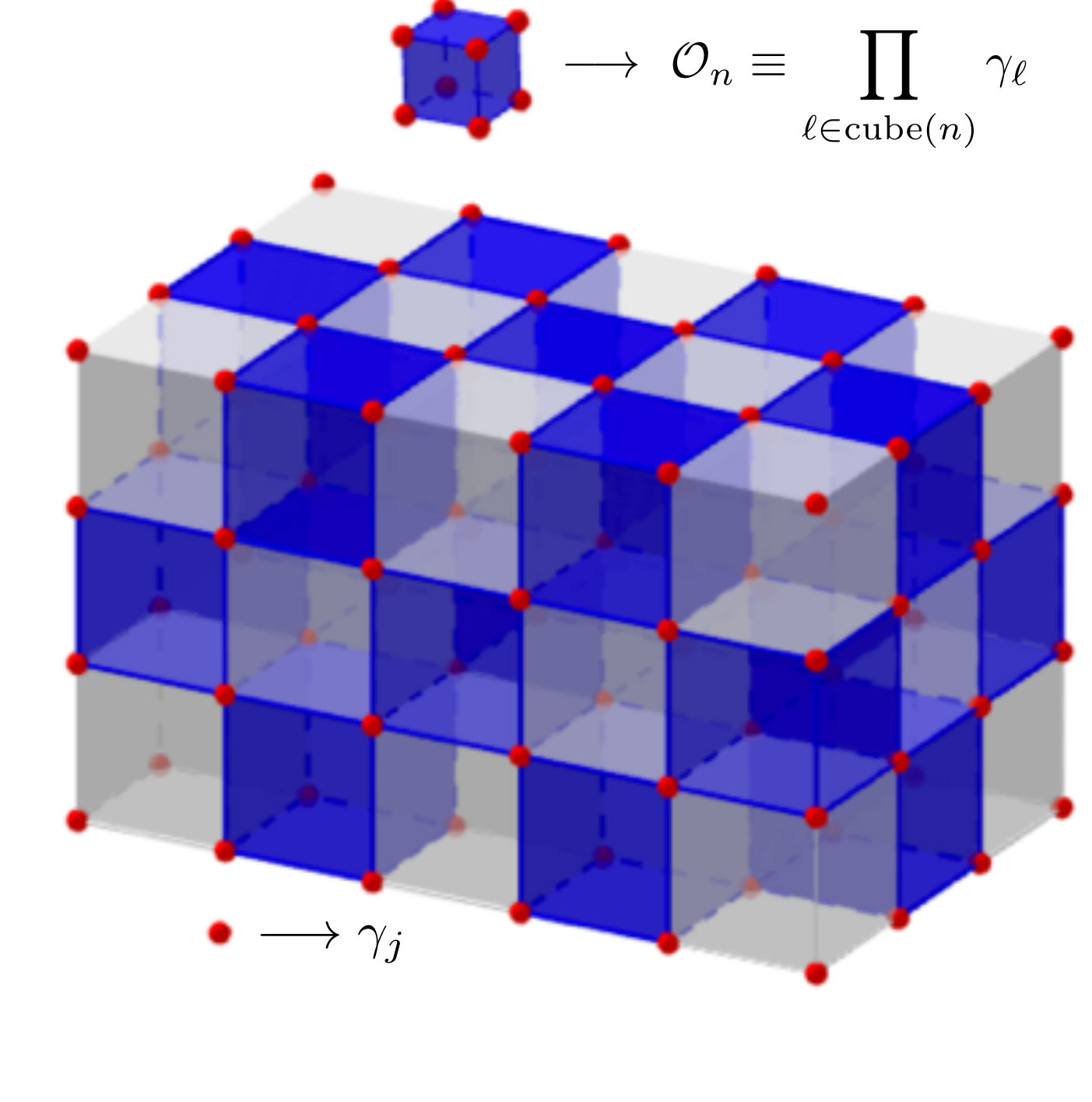}\\
 \text{(a)}\\\\
  \includegraphics[trim = 0 0 0 0, clip = true, width=0.3\textwidth, angle = 0.]{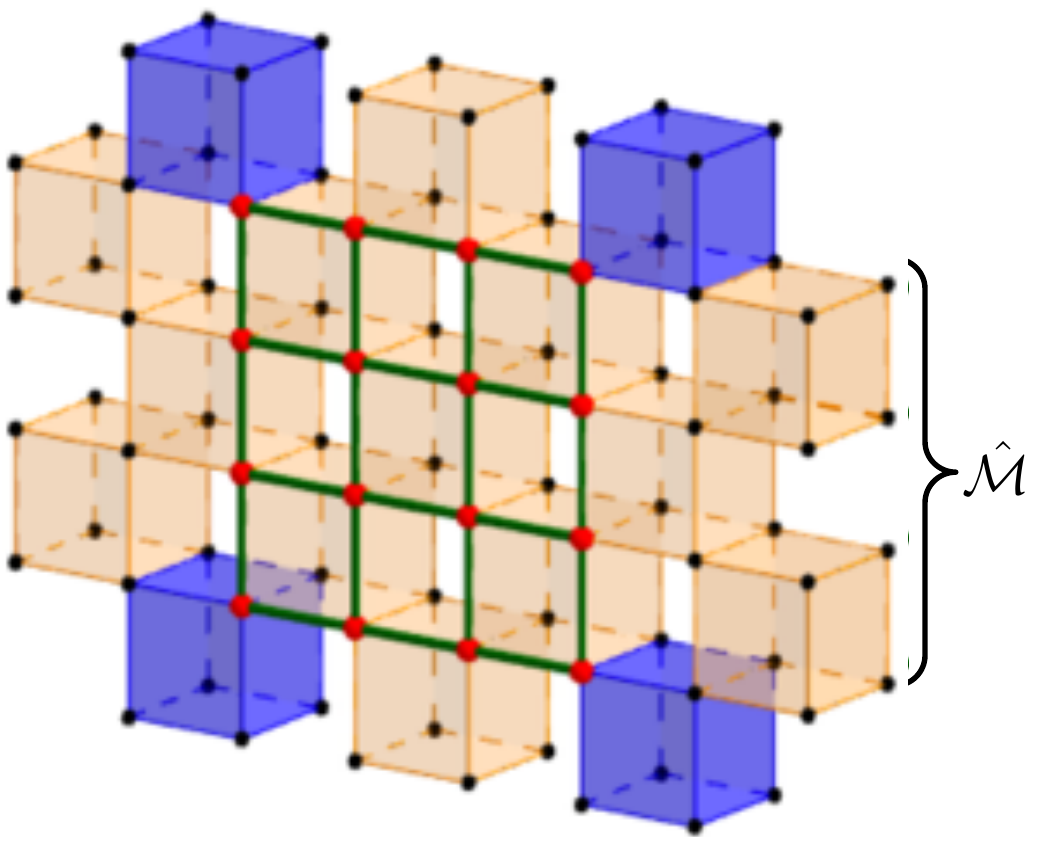}\\
\text{(b)}
  \end{array}$ 
 \caption{{\bf Checkerboard Model:} Majorana fermions are placed on the sites of a three-dimensional cubic lattice.  At each cube, the fermion parity is defined by the product of the eight Majorana fermions at the vertices.  The checkerboard model is given by summing this operator over the blue cubes on the lattice, which form a checkerboard array.  Adapted from Ref. \cite{Vijay1}.}
  \label{fig:Checkerboard}
\end{figure}

To analyze the properties of the checkerboard model, it is convenient to recall that any complex fermion $c$, $c^{\dagger}$ may be re-written in terms of a pair of Majorana fermions as 
\begin{align}
c^{\dagger} = \frac{\gamma_{1} + i\gamma_{2}}{2}\hspace{.25in} c = \frac{\gamma_{1} - i\gamma_{2}}{2}
\end{align}
where the Majorana fermions $\gamma_{i}$ satisfy canonical anti-commutation relations $\{\gamma_{i},\gamma_{j}\} = \delta_{ij}$.  Rewriting each gauge fermion as a pair of Majorana fermions yields a simple representation of the checkerboard model (\ref{Hc}) in terms Majorana fermions on the \emph{sites} of a three-dimensional cubic lattice.  The Hamiltonian for the checkerboard model (\ref{Hc}) now takes the simple form
\begin{align}\label{eq:Hcheck}
H_{c} = -\sum_{c}\,'\,\mathcal{O}_{c}
\end{align}
where $\mathcal{O}_{c}$ is defined as the product of the eight Majorana operators on the vertices of cube $c$ 
\begin{align}
\mathcal{O}_{c} = \prod_{j\in c}\gamma_{j}
\end{align}
 while the sum in (\ref{eq:Hcheck}) is taken over a checkerboard array of cubes, as indicated by the colored blue cubes in Fig. \ref{fig:Checkerboard}a.  To connect with the presentation in the main text, we note that after this re-writing, the operators $B_{p,\ell}$ that favor zero-flux of the $Z_{2}$ gauge field become cube operators on the even layers of cubes in the checkerboard array, while the inter-layer interaction $\mathcal{O}_{j,\ell}$ between gauge fermions becomes a cube operator on the odd layers. We observe that all of the operators $\mathcal{O}_{c}$ mutually commute and square to the identity 
\begin{align}
\mathcal{O}_{c}^{2} = +1 \hspace{.5in} [\mathcal{O}_{c},\,\mathcal{O}_{c'}] = 0
\end{align}
so that the ground-state of (\ref{eq:Hcheck}) satisfies $\mathcal{O}_{c}\ket{\Psi_{\mathrm{gs}}} = \ket{\Psi_{\mathrm{gs}}}$ for all of the cubes.

 \begin{figure}
 $\begin{array}{cc}
 \includegraphics[trim = 300 245 250 200, clip = true, width=0.3\textwidth, angle = 0.]{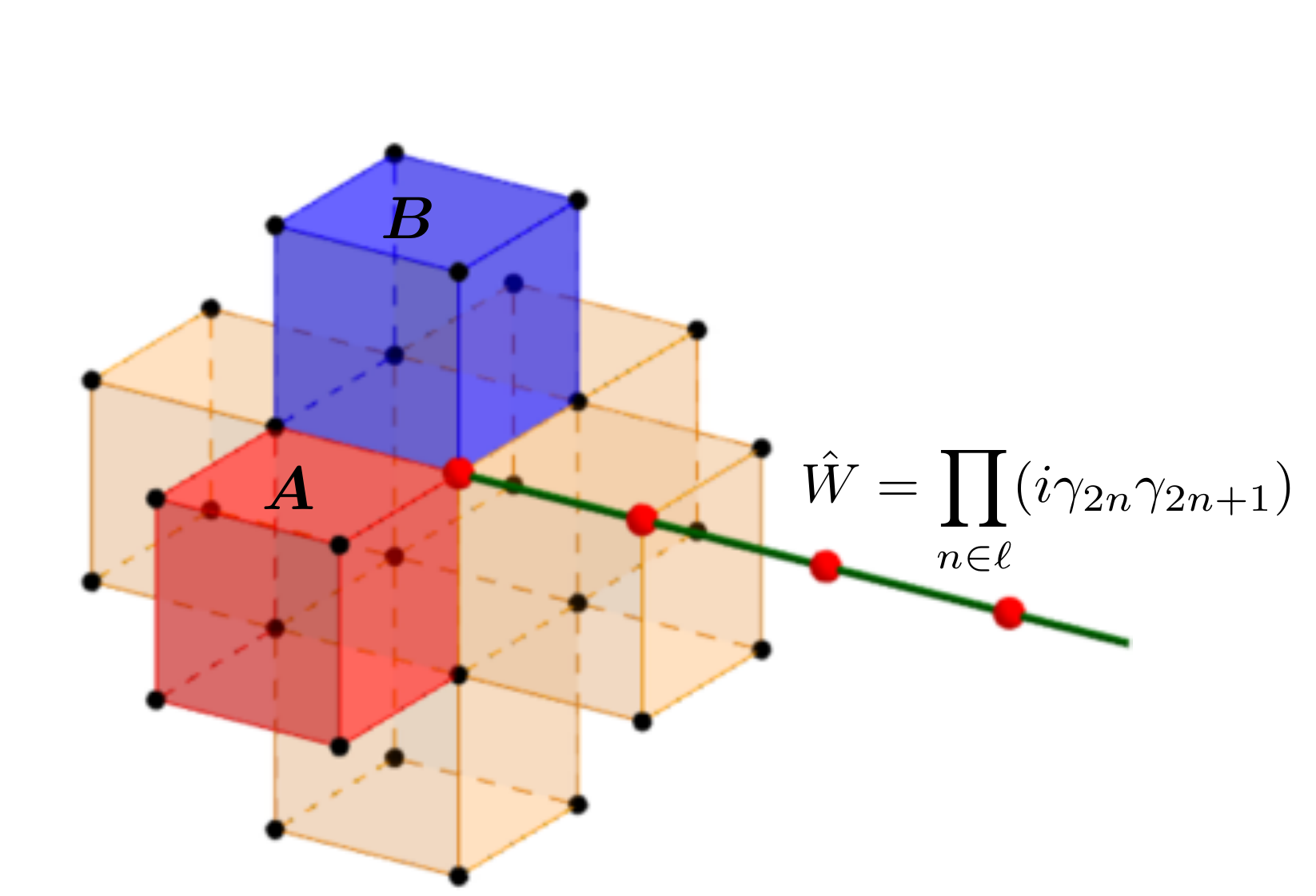} &
  \includegraphics[trim = 300 200 370 200, clip = true, width=0.18\textwidth, angle = 0.]{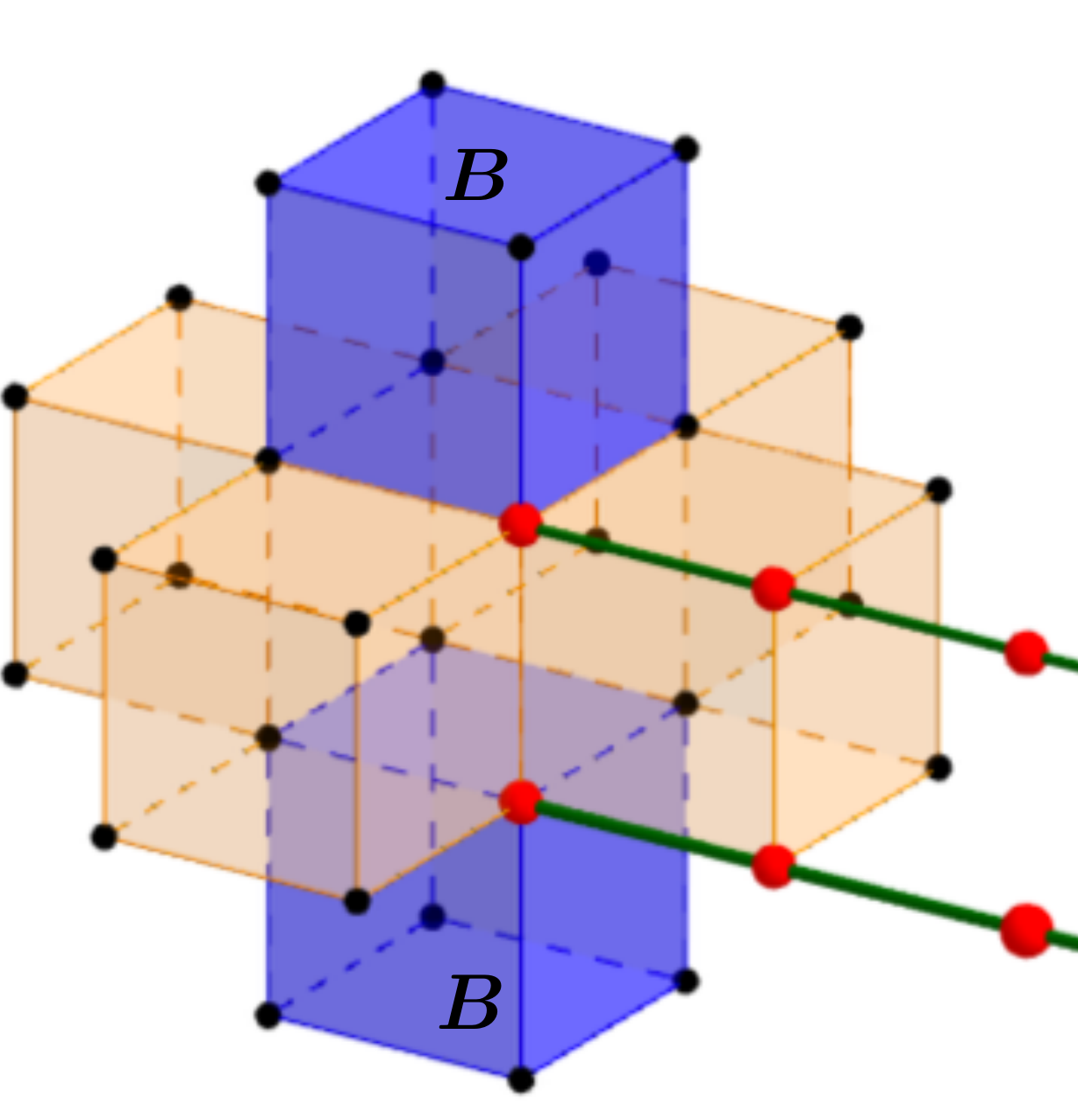}\\
\text{(a)} & \text{(b)}
  \end{array}$ 
 \caption{{\bf Mobile Fractionalized Excitations:} Pairs of fractons -- appearing (a) at the ends of a line-like operator or at the end of (b) a ``short" membrane -- are fractionalized excitations that with reduced mobility. The excitation in (a) can move along the line defined by the line-like operator while the one in (b) is free to move in a plane orthogonal to the short membrane.  Adapted from Ref. \cite{Vijay1}.}
  \label{fig:QP}
\end{figure}

As shown in Ref. \cite{Vijay1}, the Hamiltonian (\ref{eq:Hcheck}) possesses an exotic kind of three-dimensional topological order with immobile, fractionalized excitations.  First, we note that the ground-state of the Hamiltonian exhibits a sub-extensive degeneracy when placed on the three-torus, and that remarkably, the ground-states are locally indistinguishable \cite{Vijay1}. In other words, fixing the eigenvalues of $\{\mathcal{O}_{c}\}$ determines all of the local observables in the system.  Therefore, certain fundamental properties of the solvable Hamiltonian (\ref{eq:Hcheck}), including its degeneracy on the torus and the nature of the fractionalized excitations above the ground-state, are robust even in the presence of local perturbations and are essential properties of a stable phase of three-dimensional quantum matter.

The sub-extensive ``topological" degeneracy (e.g. $D = 2^{6L-6}$ on the $L\times L\times L$ three-torus \cite{Vijay1}) arises due to the fact that the gapped, fractionalized excitations that may be created in the checkerboard model have severely restricted mobility.   The fundamental excitations above the ground-state are obtained by acting with local operators to flip the eigenvalue of certain cube operators appearing in (\ref{eq:Hcheck}).  For example, acting on the ground-state with a membrane-like operator $\hat{\mathcal{M}}$ which is given by a product of Majorana operators over the sites of a flat, rectangular region of the lattice as shown in Fig. \ref{fig:Checkerboard}b, creates four excitations at the \emph{corners} of the membrane ($\mathcal{O}_{c}=-1$ for the blue cubes in Fig. \ref{fig:Checkerboard}b).  The elementary fractionalized excitation ($\mathcal{O}_{c} = -1$) can only be created at the corners of a membrane-like operator -- in stark contrast to point-like excitations in a conventional gauge theory, which appear at the ends of Wilson line operators -- due to the fact that on the torus, the product of the $\mathcal{O}_{c}$ operators along any plane ($xy$, $yz$ or $xz$) is equal to the identity.  These excitations, termed ``fractons" \cite{Vijay2}, are therefore \emph{immobile}.  That is, attempting to move a single excitation at the corner of the membrane will necessarily nucleate other gapped excitations in the system.  

Pairs of fractons, however, behave as fractionalized excitations with restricted mobility.  This is most simply seen by taking one dimension of the membrane-like operator $\hat{\mathcal{M}}$ to be small.  The resulting excitation (the ``dimension-2" quasiparticle in the language of Ref. \cite{Vijay1}) , as shown in Fig. \ref{fig:QP}b, is free to move along a plane orthogonal to the membrane, and has nontrivial mutual statistics with a fracton excitation contained within its plane of motion.  The gauge-matter coupling discussed in the main text amounts to binding the fermionic excitations of the superconductor to this dimension-2 quasiparticle. 

\section{Non-Abelian Fractons and Braiding}
\begin{figure}
 $\begin{array}{ccc}
 \includegraphics[trim = 0 0 0 0, clip = true, width=0.16\textwidth, angle = 0.]{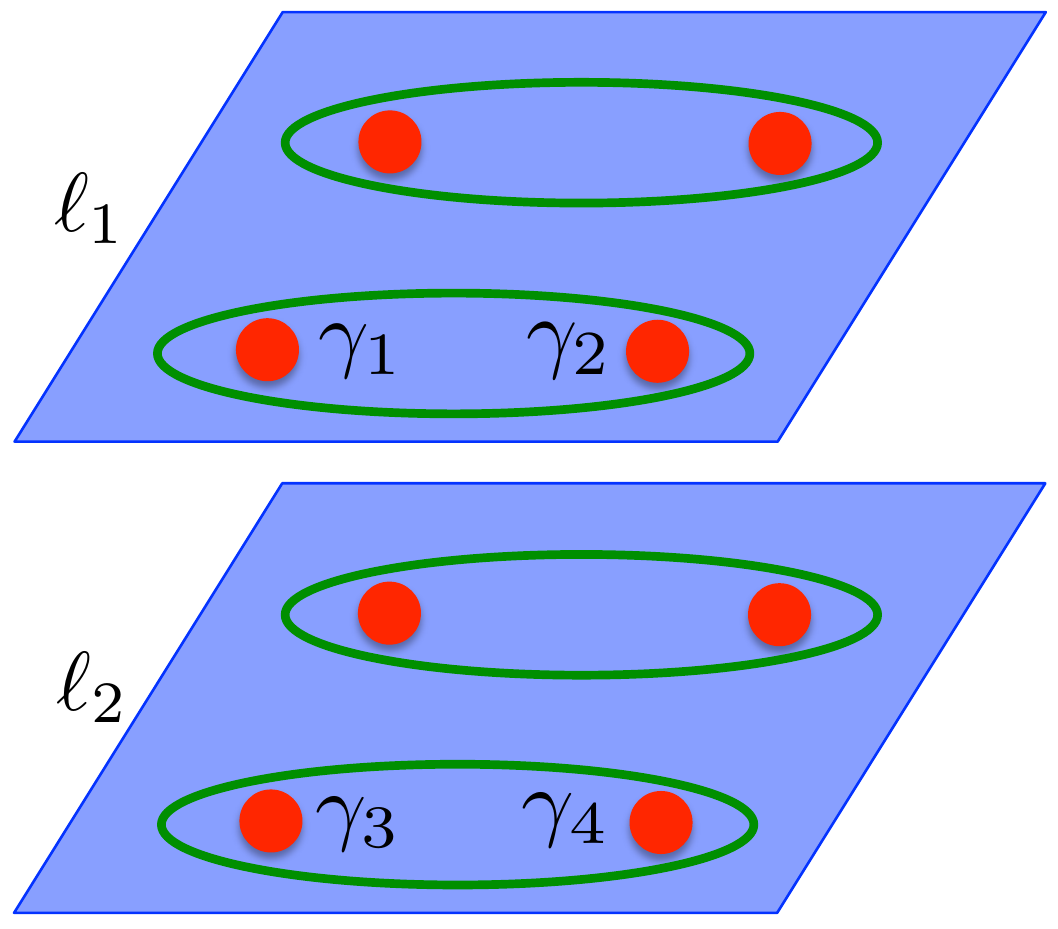}
 & \includegraphics[trim = 0 0 0 0, clip = true, width=0.16\textwidth, angle = 0.]{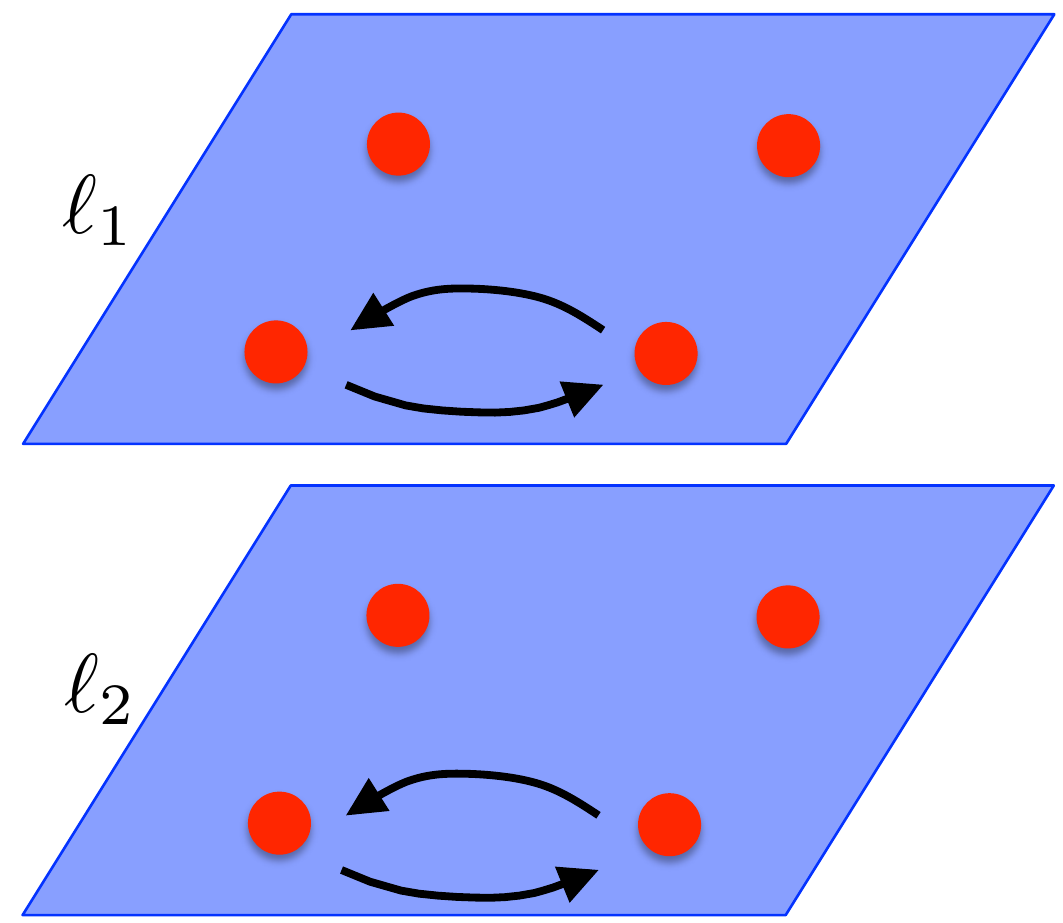}\,\, &
 \,\, \includegraphics[trim = 0 0 0 0, clip = true, width=0.16\textwidth, angle = 0.]{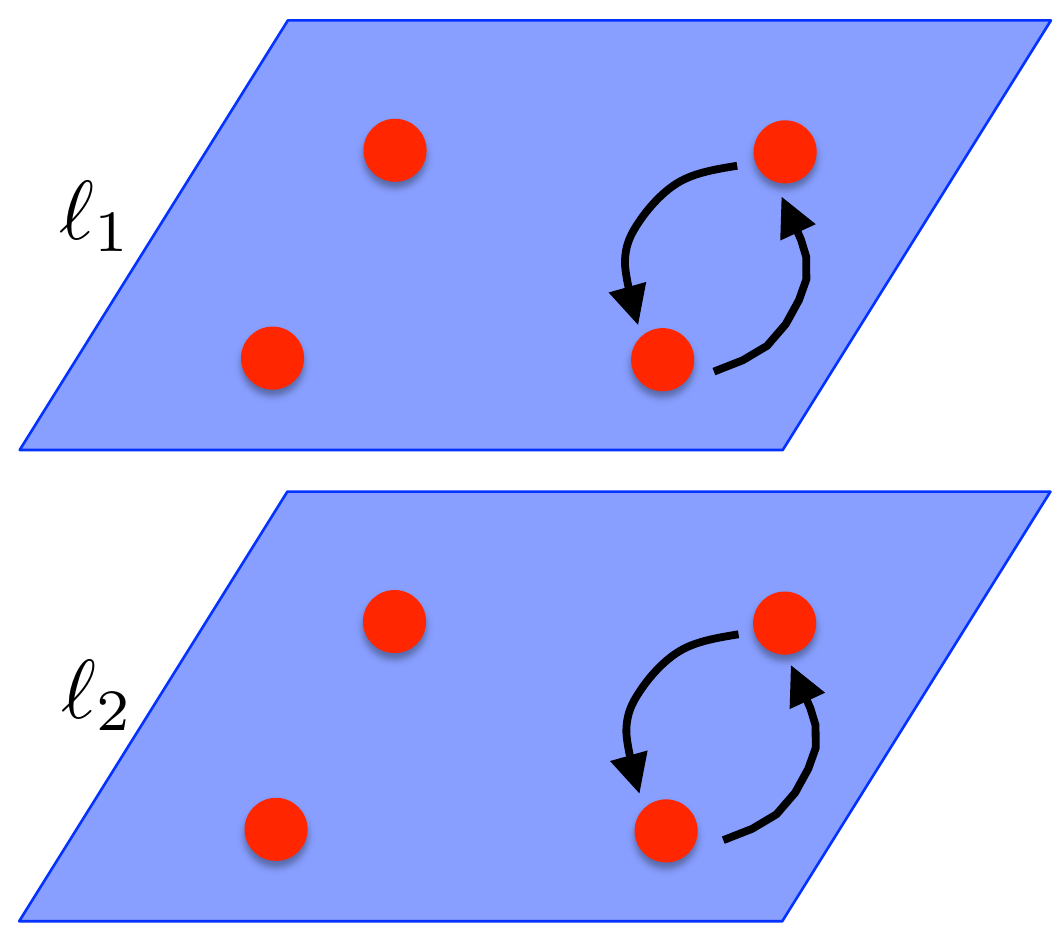}\\
 & & \\
 \text{(a)} & \text{(b)} & \text{(c)}
 \end{array}$
 \caption{{\bf Braiding Transformations:} By applying membrane-like operators, we may exchange pairs of fractons in nearby layers to affect unitary transformations on the degenerate states in the Hilbert space.   }
  \label{fig:Braiding_Transf}
\end{figure}

In this section, we demonstrate that appropriately moving and braiding the non-Abelian anyon formed from a pair of fractons ($\sigma_{\ell}\times\sigma_{\ell'}$) in our model, implements a unitary transformation on the space of degenerate, locally indistinguishable states. 
The nature of the unitary transformation may be determined, up to an overall Abelian phase, by observing that a membrane-like operator that moves a pair of fractons in distinct layers must conserve the parity of the matter fermions $U^{(f)}_{\ell}$ within each layer, as required by gauge invariance, and so that no other excitations are created in the system.  Therefore, a vertical membrane operator that exchanges pairs of well-separated fractons, as in Fig. \ref{fig:Braiding_Transf}b, must affect the following transformation on the fermionic zero mode operators shown $\gamma_{1}\rightarrow \gamma_{2}$, $\gamma_{2}\rightarrow -\gamma_{1}$, $\gamma_{3}\rightarrow \gamma_{4}$, $\gamma_{4}\rightarrow -\gamma_{3}$, up to an overall choice of sign in each layer.  

The resulting unitary transformation may be written by pairing the fractons as shown in Fig. \ref{fig:Braiding_Transf}a, to form a basis for the protected, $2^{4 - 2} = 4$-dimensional Hilbert space shown.  Using basis states $\ket{1,1}$, $\ket{1, \psi}$, $\ket{\psi, 1}$, $\ket{\psi, \psi}$ which  describe the fusion channels of a pair of fractons in layer $\ell_{1}$ and in $\ell_{2}$, respectively, and using the $F$- and $R$- matrices for Ising anyons \cite{Kitaev},  we observe that the fracton exchange shown in Fig. \ref{fig:Braiding_Transf}b affects the diagonal transformation $B_{1} = e^{i\theta_{\Gamma}}\,\mathrm{diag}\left(e^{-i\pi/4},\,e^{i\pi/4},\,e^{i\pi/4},\,-e^{-i\pi/4}\right)$, while the exchange process in Fig. \ref{fig:Braiding_Transf}c is off-diagonal
\begin{align}
B_{2} = e^{i\theta_{\Gamma}}\,B\otimes B.
\end{align}
where $B \equiv i{e^{i\pi/4}}{(1 -i \tau^{x})}/{2}$ and $\tau^{x}$ is the spin-1/2 Pauli-$X$ operator. Here, the overall Abelian Berry phase $e^{i\theta_{\Gamma}}$ can depend on the number of fractons that lie in the planes \emph{between} $\ell_{1}$ and $\ell_{2}$, and that have been enclosed by the membrane operator used to exchange the fractons.  Such a phase arises since pairs of fractons in the Majorana checkerboard model have $\pi$ mutual statistics with fractons contained within their plane of motion.

\section{Coupled $G$ Gauge Theories and Non-Abelian Fractons}
\begin{figure}
 \includegraphics[trim = 0 0 0 0, clip = true, width=0.2\textwidth, angle = 0.]{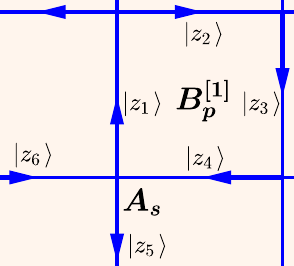}
 \caption{{\bf Quantum Double Layer:} A single layer of the quantum double model \cite{Quantum_Double} on the square lattice.  The action of the star ($A_{s}$) and plaquette ($B_{p}$) operators on each link is described in the text.}
  \label{fig:Quantum_Double}
\end{figure}

Consider the quantum double model, as originally introduced in Ref. \cite{Quantum_Double}, which describes the zero-correlation length limit of the deconfined phase of a 2D gauge theory with finite gauge group $G$.  Within each layer, degrees of freedom are placed on the oriented links of a square lattice, and labeled by the elements of the finite group $G$.  The Hamiltonian takes the form
\begin{align}
H_{\ell} = -\sum_{s}A_{s} - \sum_{p}B_{p}^{[1]}
\end{align}
where the ``flux" operator at a plaquette $p$ projects onto a state where the oriented product of the degrees of freedom along links surrounding $p$ are equal to the identity. We define $B_{p}^{[g]}$ as
\begin{align}
B_{p}^{[g]} \equiv \sum_{z_{1}z_{2}z_{3}z_{4} \in [g]} \ket{z_{1},z_{2},z_{3},z_{4}}\bra{z_{1},z_{2},z_{3},z_{4}}
\end{align}
where $[g]$ is the conjugacy class associated with the group element $g\in G$.  The ``star" operator $A_{s}$ multiplies the oriented links surrounding site $s$ by elements of $G$ as follows:
\begin{align}
A_{s} &\equiv \frac{1}{|G|}\sum_{g\in G} A_{s}^{(g)}
\end{align}
where
\begin{align}
A_{s}^{(g)} &\equiv \sum_{\{z_{i}\} \in G} \ket{gz_{1}, z_{4}g^{-1}, gz_{5}, z_{6}g^{-1}}\bra{z_{1}, z_{4}, z_{5}, z_{6}} + \mathrm{h.c.}\nonumber
\end{align}
with $z_{1}$, $\ldots$, $z_{6}$ as shown in Fig. \ref{fig:Quantum_Double}.  The two operators $A_{s}$, $B_{p}^{[g]}$ commute and are both projection operators.  As a result, the ground-state of a single layer satisfies $A_{s}\ket{\Psi} = B_{p}^{[1]}\ket{\Psi} = \ket{\Psi}$.

The quantum double model, which describes the zero correlation-length limit of all lattice gauge theories with a finite gauge group in (2+1)-dimensions, admits gapped charge and flux excitations, which live on the sites and plaquettes of the lattice, respectively, and are obtained by acting on the ground-state with a ``ribbon"-like operator \cite{Quantum_Double}.   Pure charge excitations are labeled by irreducible representations of $G$, while pure fluxes are labeled by conjugacy classes $\mathcal{C}$ of the group.  More generally, a charge-flux composite (dyon) is labeled by $\mathcal{C}$, as well as an irreducible representation of the \emph{centralizer} of $\mathcal{C}$, denoted C$_{\mathcal{C}}$.  As an example, $G = S_{3}$ -- the permutation group on three elements -- may be parametrized as
\begin{align}
S_{3} = \{1, x, y, y^{2}, xy, yx\}
\end{align}
where the elements $x$ and $y$ satisfy
\begin{align}
x^{2} = 1\hspace{.4in} y^{3} = 1\hspace{.4in} yx = xy^{2}
\end{align}
The three conjugacy classes of $S_{3}$ -- $\{1\}$, $\{y,\,y^{2}\}$, and $\{x, xy, yx\}$, which we label as $[1]$, $[x]$ and $[y]$ respectively -- have centralizers $S_{3}$, $\mathbb{Z}_{2}$ and $\mathbb{Z}_{3}$, respectively.  The irreducible representations of $S_{3}$ include the trivial representation, the sign representation (``sgn"), as well as the standard two-dimensional representation.  A full list of the eight excitations in the $S_{3}$ lattice gauge theory are given below
\begin{align*}
  \begin{tabular}{|c||c||c||c|}
  \hline
    \,\,Excitation\,\, & \,\,Flux\,\, & \,\,Charge\,\, & $\,\,d\,\,$\\
    \hline
    {\bf A} & $\cdot$ & $\cdot$ & 1\\
    \hline
    {\bR B} & $\cdot$ & sgn & 1\\
    \hline
     {\bR C} & $\cdot$ & 2 & 2\\
    \hline
    {\bB D} & $[x]$ & $\cdot$ & 3\\
    \hline
     {\bG E} & $[x]$ & $-1$ & 3\\
    \hline
    {\bB F} & $[y]$ & $\cdot$ & 2\\
    \hline
    {\bG G} & $[y]$ & $\omega$ & 2\\
    \hline
    {\bG H} & $[y]$ & $\overline{\omega}$ & 2\\
    \hline
    \end{tabular}
\end{align*}
Here $\omega$ and $\overline{\omega}$ label the two non-trivial representations of $\mathbb{Z}_{3}$ while $-1$ denotes the non-trivial representation of $\mathbb{Z}_{2}$.  The last column is the quantum dimension of each excitation, which is given by
\begin{align}
d_{\mathcal{C}, \mathrm{Rep}(\mathrm{C}_{\mathcal{C}})} \equiv |\mathcal{C}|\cdot\mathrm{dim}\,\mathrm{Rep}(\text{C}_{\mathcal{C}}).
\end{align}
The coloring of the excitations is to denote the neutral vacuum (black), pure charge (red), flux (blue) or a dyon (green) excitation.  For our purposes, we will be concerned with the non-Abelian flux {\bB D}, whose relevant fusion rules for our purposes we write suggestively as ${\bB D}\times {\bf A} = {\bB D}$;  ${\bB D}\times {\bB F} = {\bB D}\times ( {\bf 1} + {\bR B})$; ${\bB D}\times {\bB D} = ({\bf 1} + {\bR C}) + {\bB F}\times ({\bf 1} + {\bR C})$.

We now consider intersecting layers of the quantum double model for a group $G$, where each layer $\ell$  is described by the Hamiltonian
\begin{align}
H_{\ell} = -\sum_{s}A_{s,\ell} - \sum_{p}B_{p,\ell}^{[1]}
\end{align}
The layers are placed in an intersecting configuration, as shown in Fig. \ref{fig:Quantum_Double_Layers}a of the main text, to form a cubic lattice with two $G$-degrees of freedom per link.  Furthermore, the layers are arranged so that a pair of overlapping links from orthogonal layers have opposite orientation, as shown in Fig. \ref{fig:Quantum_Double_Layers}a. The state on link $\langle s,s'\rangle$ is now labeled $\ket{z_{ss'}, w_{ss'}}$.  On a given link, we may explicitly write the operator that creates two pure flux excitations in each of the adjacent, orthogonal layers meeting at link $\langle s, s'\rangle$ as $W^{[g]}_{{ss'}}\otimes \widetilde{W}_{{ss'}}^{[g]}$ where
\begin{align}
W_{{ss'}}^{[g]} \equiv \frac{1}{|G|}\sum_{z_{ss'}\in G}\sum_{h\in[g]} \ket{z_{ss'}h}\bra{z_{ss'}}
\end{align}
acts on $z_{ss'}$, while $\widetilde{W}_{{ss'}}^{[g]}$ acts identically on other degree of freedom ($w_{ss'}$) on the same link. We claim that the following Hamiltonian for the coupled quantum double layers
\begin{align}
{H = \sum_{\ell}H_{\ell} - \Delta\sum_{\langle s,\,s'\rangle}\mathcal{O}_{ss'}^{[g]}}
\end{align}
where
\begin{align}
\mathcal{O}_{ss'}^{[g]}  = W^{[g]}_{{ss'}}\otimes \widetilde{W}_{{ss'}}^{[g]} + \mathrm{h.c.}
\end{align}
can give rise to a fracton topological phase in the limit that $\Delta \gg 1$, for an appropriate choice of flux $[g]$.  The action of $\mathcal{O}_{ss'}^{[g]}$ on the decoupled layers of quantum double models is shown schematically in Fig. \ref{fig:Quantum_Double_Layers}b of the main text. We refer to this process for generating a fracton topological phase, where the immobile excitations carry a non-trivial quantum dimension, as composite $[g]$ flux loop condensation.

We now study the emergence of a fracton phase from an intersecting array of quantum double layers for the group $G = S_{3}$ -- the permutation group on three elements -- after condensing the composite flux loop formed from the non-Abelian flux {\bB D} associated with the conjugacy class $[x]$. When $\Delta \gg 1$, the state on each link must satisfy 
\begin{align}\label{eq:constraint}
\mathcal{O}_{ss'}^{[x]}\ket{\widetilde{h_{ss'}}} = \frac{1}{2}\ket{\widetilde{h_{ss'}}}
\end{align}
and we may derive a low-energy effective theory that acts exclusively within this subspace. 
 \begin{figure}
 \includegraphics[trim = 0 0 0 0, clip = true, width=0.2\textwidth, angle = 0.]{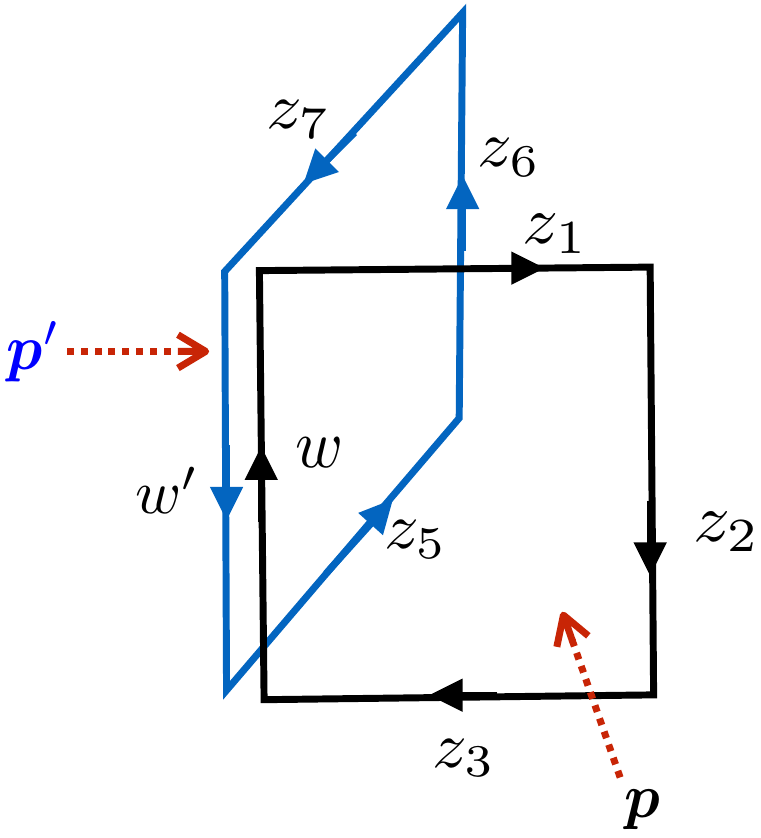}
 \caption{Setup for computing the commutator of $\mathcal{O}^{[g]}_{ss'}$ and $O_{c}$, as discussed in the text.}
  \label{fig:Commutation}
\end{figure}
To derive this effective Hamiltonian, we begin by noting that $[A_{s,\ell},\mathcal{O}_{ss'}^{[x]}] = 0$, which may be verified by explicit calculation.  Additionally, observe that the flux operators $B^{[g]}_{p,\ell}$ do not generally commute with $\mathcal{O}_{ss'}^{[x]}$, since the latter has the effect of creating {\bB D} flux excitations.   Specific products of the charge and flux operators may commute with $\mathcal{O}_{ss'}^{[x]}$, however, as certain local conservation laws may remain even after condensing the non-Abelian {\bB D} flux loop.  When $\Delta \gg 1$, we find that the effective Hamiltonian takes the form
\begin{align}\label{eq:H_eff}
H_{\mathrm{eff}} = -\sum_{s,\ell}A_{s,\ell} - K\sum_{c}O_{c} - \cdots
\end{align}
where $K \sim \Delta^{-6}$ is determined in perturbation theory, and the operator $O_{c}$ is defined as
\begin{align}
O_{c} \equiv \prod_{p,\ell \in\partial c}\left(2B_{p,\ell}^{[x]}-1\right) 
\end{align}
with the product taken over the six plaquettes surrounding cube $c$.   The ellipsis in (\ref{eq:H_eff}) denotes other operators that commute with $\mathcal{O}_{ss'}^{[x]}$ at all sites that arise from higher-order perturbation theory.  

The second term appearing in the effective Hamiltonian reflects the fact that after condensing the composite {\bB D} flux loop, any configuration of fluxes appearing in the ground-state must be such that there are an even number of these fluxes at every cube on the lattice.  The zero-flux condition in a single quantum double layer has been replaced by an emergent constraint on the parity of the {\bB D} fluxes at each cube.  We now verify that this is indeed an emergent conservation law (i.e. $[O_{c}, \mathcal{O}_{ss'}^{[x]}] = 0$) by explicitly by evaluating the commutator between the operators $(2B_{p,\ell}^{[x]}-1) (2B_{p',\ell'}^{[x]}-1)$ and $\mathcal{O}^{[x]}_{ss'}$ -- as defined on the orthogonal plaquettes $p$ and $p'$ that overlap on the link $\langle s, s'\rangle$ as shown in Fig. \ref{fig:Commutation} -- when acting on an arbitrary state $\ket{\psi} \equiv \ket{z_{1},z_{2},z_{3},z_{4},z_{5},z_{6},w,w'}$.  Observe that
\begin{align}
&\left[\mathcal{O}^{[x]}_{ss'},\,(1 - 2B_{p,\ell}^{[x]}) (1 - 2B_{p',\ell'}^{[x]})\right]\ket{\psi}\nonumber\\
&= \frac{1}{|S_{3}|}\sum_{h, h'\in[x]} f(\{z_{i}\},w,w',h,h')\,\ket{\{z_{i}\},w h,w' h'}\nonumber
\end{align}
where 
\begin{align}
f(\{z_{i}\},w,w', h, &h') = (1 - 2\,\delta_{z_{6}z_{7}w'z_{5}\in[x]})(1-2\,\delta_{w z_{1}z_{2}z_{3}\in[x]})\nonumber \\
&- (1 - 2\,\delta_{z_{6}z_{7}w'h'z_{5}\in[x]})(1-2\,\delta_{w h z_{1}z_{2}z_{3}\in[x]})\nonumber
\end{align}
Here the Kronecker delta $\delta_{g\in[x]} = 1$ iff $g\in[x]$ and is $0$ otherwise. We observe that $a\cdot b\in[x]$ iff only one of the elements $a$ or $b$ is a member of $[x]$.  Therefore, for any $g$, $k\in S_{3}$ and $h\in[x]$, we find that $(1 - 2\delta_{ghk\in[x]}) = -(1 - 2\delta_{gk\in[x]})$.  This immediately implies that $f(\{z_{i}\},w,w',h,h') = 0$ and confirms the claim that $[O_{c},\,\mathcal{O}^{[x]}_{ss'}]=0$.

To summarize, we have shown that even after condensing the composite {\bB D} flux loop, the parity of the fluxes at every cube on the lattice remains well-defined.  While this parity is fixed to be $O_{c}= +1$ in the ground-state, acting with line- and membrane-like operators can create patterns of gapped excitations ($O_{c} = -1$) which are fundamentally immobile.  Observe that when $\Delta \gg 1$, acting on the ground-state with a Wilson line \cite{Quantum_Double} will create four such excitations, by anti-commuting with four of the cube operators $O_{c}$; each pair of excitations will be located at the two ends of the Wilson line.  As explained in the main text, an array of these Wilson lines will create four such excitations that are well-separated and which are fundamentally immobile due the geometry of the membrane-like operator, i.e. a single excitation cannot be moved without creating other such excitations in the system.  As in the Majorana checkerboard model, the product of the $O_{c}$ operators along any plane is equal to the identity, after imposing periodic boundary conditions.  This non-local constraint implies that the cube excitations may only be created in clusters of four at the corners of an operator with support on a flat, membrane-like region.  As advertised, the non-Abelian {\bB D} flux remains deconfined after coupling the $S_{3}$ quantum double layers, and has become an immobile fracton excitation in the condensed phase.   

Other excitations also remain deconfined in the condensed phase.  We do not present an exhaustive list of these excitations here, but instead argue that when $\Delta \gg 1$ a bound pair of the {\bR B} charges in orthogonal layers, which is only free to move along a line without proliferating additional excitations (i.e. a ``dimension-1 quasiparticle" in the language of Ref. \cite{Vijay2}), remains a deconfined excitation with reduced mobility, while all of the pure charge excitations are confined.   Recall that within a single layer of the 2D $G$ gauge theory, the operator that creates a pure charge excitation, associated with the irreducible representation $R$ of the group $G$ is given by
\begin{align}\label{eq:W_c}
\mathcal{W}_{ss'}^{(R)} = \sum_{z_{ss'}\in G}\chi_{_{R}}(z_{ss'})\ket{z_{ss'}}\bra{z_{ss'}}
\end{align}
where $\chi_{_{R}}(g) = \mathrm{Tr}_{_{R}}(g)$ is a character of the representation $R$.  The operator that creates the non-Abelian {\bB D} flux does not commute with (\ref{eq:W_c}).  Observe that $[W_{ss'}^{[x]}, \mathcal{W}_{ss'}^{(R)}] \ne 0$, so that an isolated charge cannot be created within the low-energy subspace defined by (\ref{eq:constraint}) when $\Delta \gg 1$. Notably, however, the operator that creates two {\bR B} charges in orthogonal layers at a given link commutes with the composite {\bB D} flux loop condensation, i.e. $[\mathcal{W}_{ss'}^{(\mathrm{sgn})}\otimes\widetilde{\mathcal{W}}_{ss'}^{(\mathrm{sgn})}, \mathcal{O}_{ss'}^{[x]}] = 0$, as may be verified by explicit calculation, using the fact that the characters for the sign representation of $S_{3}$ are simply $\chi_{_{\mathrm{sgn}}}(1)=\chi_{_{\mathrm{sgn}}}(y)=1$, $\chi_{_{\mathrm{sgn}}}(x)=-1$. Here, the operators $\mathcal{W}_{ss'}^{(\mathrm{sgn})}$ and $\widetilde{\mathcal{W}}_{ss'}^{(\mathrm{sgn})}$ act on the two degrees of freedom $z$ and $w$ at link $\langle s, s'\rangle$, respectively:
\begin{align}\label{eq:dim1qp}
\mathcal{W}_{ss'}^{(R)}\otimes\widetilde{\mathcal{W}}_{ss'}^{(R)} \equiv \sum_{z,w\in G}\chi_{_{R}}(z_{ss'})\chi_{_{R}}(w_{ss'})\ket{z_{ss'},w_{ss'}}\bra{z_{ss'},w_{ss'}}\nonumber
\end{align}
Therefore, we conclude that while all of the isolated pure charge excitations are confined, a bound pair of {\bR B} charges in orthogonal layers is a deconfined excitation, which may be moved along a line by sequentially applying the operator $\mathcal{W}_{ss'}^{(\mathrm{sgn})}\otimes\widetilde{\mathcal{W}}_{ss'}^{(\mathrm{sgn})}$.  Since this excitation is formed from a composite of excitations in adjacent layers, it is only free to move along the line along which the layers intersect, and is therefore a dimension-1 quasiparticle in the language of Ref. \cite{Vijay1}.    
\end{document}